\documentclass[amstex,aps,preprint,showpacs]{revtex4}%
\usepackage{graphicx}
\usepackage{amsmath}
\usepackage{bm}%
\usepackage{amsfonts}%
\usepackage{amssymb}

\begin{document}

\title{Classical statistical distributions can violate Bell-type inequalities}
\author{A.\ Matzkin}
\affiliation{Laboratoire de Spectrom\'{e}trie Physique (CNRS Unit\'{e} 5588),
Universit\'{e} Joseph-Fourier Grenoble-1, BP 87, 38402 Saint-Martin
d'H\`{e}res, France}

\begin{abstract}
We investigate two-particle phase-space distributions in classical
mechanics constructed to be the analogs of quantum mechanical
 angular momentum eigenstates. We obtain the phase-space averages of
 specific observables related to the
projection of the particles' angular momentum along axes with
different orientations, and show that the ensuing correlation
function violates Bell's inequality. The key to the violation
resides in choosing observables impeding the realization of the
joint measurements whose existence is required in the derivation of
the inequalities. This situation can have statistical (detection
related) or dynamical (interaction related) underpinnings, but
non-locality does not play any role.

\end{abstract}


\maketitle

\section{Introduction}

Bell's theorem was originally introduced \cite{bell1964} to examine
quantitatively the consequences of postulating hidden variable distributions
on the incompleteness of quantum mechanics put forward by Einstein, Podolsky
and Rosen \cite{EPR} (EPR). The core of the theorem takes the form of
inequalities involving average values of two observables each related to one
of the two particles. Bell showed that these inequalities must be satisfied by
any theory containing local variables aiming to complete quantum mechanics in
the EPR sense. The assumptions leading to Bell's theorem imply the existence
of a joint probability distribution accounting for the simultaneous existence
of incompatible quantum observables \cite{fine82,cabello96}. Local models
forbidding the existence of these joint distributions are therefore not bound
by Bell's theorem. Indeed local models violating Bell-type inequalities have
already been proposed \cite{beltrametti,orlov,christian}, but these models are
mathematical and abstract. In this work, we show that the familiar statistical
distributions of classical mechanics may lead to a violation of the relevant
Bell-type inequalities. Our main ingredients will consist first in choosing
specific classical phase-space ensembles for 2 particles (distributions
constructed to be the classical analogs of the quantum-mechanical angular
momenta eigenstates), and then in choosing detectors impeding the existence of
the joint probability distribution. We will consider two types of settings
involving angular momentum measurements, each setting being closely related to
a well-known quantum mechanical context. The first setting will consist in a
classical version of the detection loophole \cite{percival}: the relevant Bell
inequalities will be seen to be violated when the sampling is done on a
sub-ensemble, defined by the type of detected events, leading to averages
computed on a partial region of phase-space over which the joint probability
distribution cannot be defined. Hence the violation has a statistical
underpinning -- we will show there is no violation if the averages are taken
on the entire phase-space. The second setting will reveal a genuine violation
of the Bell inequalities due to dynamical reasons: by including in the
detection process a local probabilistic interaction between the measured
particle and the detector inducing a random perturbation that blurs the
particles' phase-space positions, the derivation of Bell's theorem is
effectively blocked, as only correlations between ensembles corresponding to a
fixed setting of the detectors can be made. This example can be seen as a
classical version of the quantum measurement of non-commuting observables.

\section{Classical ensembles}

We first introduce the classical analogs of the quantum mechanical
angular-momentum eigenstates to be employed below. The classical distributions
of particles can be considered either in phase-space or in configuration
space; equivalently, one can also consider the distribution of the angular
momenta on the angular momentum sphere. Let us first take a single classical
particle and assume the modulus $J$ of its angular momentum is fixed. The
value of $\mathbf{J}$ then depends on the position of the particle in the
phase-space defined by $\Omega=\{\theta,\phi,p_{\theta},p_{\phi}\},$ where
$\theta$ and $\phi$ refer to the\ polar and azimuthal angles in spherical
coordinates and $p_{\theta}$ and $p_{\phi}$ are the conjugate canonical
momenta.\ Let $\rho_{z}(\Omega)$ be the distribution in phase-space given by%
\begin{equation}
\rho_{z_{0}}(\theta,\phi,p_{\theta},p_{\phi})=N\delta(J_{_{z}}(\Omega
)-J_{z_{0}})\delta(J^{2}(\Omega)-J_{0}^{2}).\label{5}%
\end{equation}
$\rho_{z_{0}}$ defines a distribution in which every particle has an angular
momentum with the same magnitude, namely $J_{0}$, and the same projection on
the $z$ axis $J_{z_{0}}$. Hence $\rho_{z_{0}}$ can be considered as a
classical analog of the quantum mechanical density matrix $\left\vert
jm\right\rangle \left\langle jm\right\vert $ since just like a quantum
measurement of the magnitude $j$ and of the $z$ axis projection $m$ of the
angular momentum in such a state will invariably yield the eigenvalues of the
operators $\hat{J}^{2}$ and $\hat{J}_{z}$, the classical measurement of these
quantities when the phase-space distribution is known to be $\rho_{z}$ will
give $J_{0}^{2}$ and $J_{z_{0}}$ (see Appendix A). Eq. (\ref{5}) can be
integrated over the conjugate momenta to yield the \emph{configuration space}
distribution
\begin{equation}
\rho(\theta,\phi)=N\left[  \sin(\theta)\sqrt{J_{0}^{2}-J_{z_{0}}^{2}/\sin
^{2}(\theta)}\right]  ^{-1}\label{7}%
\end{equation}
where we have used the defining relations $J_{_{z}}(\Omega)=p_{\phi}$ and
$J^{2}(\Omega)=p_{\theta}^{2}+p_{\phi}^{2}/\sin^{2}\theta$. Further
integrating over $\theta$ and $\phi$ and requiring the phase-space integration
of $\rho$ to be unity allows to set the normalization constant $N=J_{0}%
/2\pi^{2}$.\

There is of course nothing special about the $z$ axis and we can define a
distribution by fixing the projection $J_{a}$ of the angular momentum on an
arbitrary axis $a$ to be constant (in this paper we will take all the axes to
lie in the $zy$ plane). Computing the distribution $\rho_{a_{0}}=\delta
(J_{a}-J_{a_{0}})\delta(J-J_{0}^{2})$ is tantamount to rotating the
coordinates towards the $a$ axis in Eq. (\ref{7}). Fig.\ 1 shows examples of
configuration space particle distributions and gives for one plot the
corresponding quantum mechanical angular-momentum eigenstate (the similarity
is not accidental, as Eq. (\ref{7}) is essentially the amplitude of the
spherical harmonic in the semiclassical regime, see Appendix A).\ We can also
determine the average projection $J_{a}$ on the $a$ axis for a distribution of
the type (\ref{7}) corresponding to a well defined value of $J_{z}$:%

\begin{equation}
\left\langle J_{a}\right\rangle _{J_{z_{0}}}=\int p_{\phi}\cos\theta_{a}%
\delta(J_{_{z}}(\Omega)-J_{z_{0}})d\Omega=J_{z_{0}}\cos\theta_{a}, \label{6}%
\end{equation}
where $\theta_{a}$ is the angle $(\widehat{z,a})$ and the projection of the
component of $J_{a}$ on the $y$ axis vanishes given the axial symmetry of the distribution.

The original derivation of the inequalities by Bell \cite{bell1964} involved
the measurement of the angular momentum of 2 spin-$1/2$ particles along
different axes. Here we will consider the fragmentation of an initial particle
with a total angular momentum $\mathbf{J}_{T}=0$ into 2 particles carrying
angular momenta $\mathbf{J}_{1}$ and $\mathbf{J}_{2}$ (we will assume to be
dealing with orbital angular momenta). Conservation of the total angular
momentum imposes $\mathbf{J}_{1}=-\mathbf{J}_{2}$ and $J_{1}=J_{2}\equiv J.$
Quantum mechanically, this situation would correspond to the system being in
the singlet state arising from the composition of the angular momenta
($j_{T}=0$, $m_{1}=-m_{2}$). Classically the system is represented by the
2-particle phase space distribution%
\begin{equation}
\rho(\Omega_{1},\Omega_{2})=N\delta(\mathbf{J}_{1}+\mathbf{J}_{2}), \label{9}%
\end{equation}
where $N$ is again a normalization constant. On the angular momentum sphere
the distribution (\ref{9}) corresponds to $\mathbf{J}_{1}$ and $\mathbf{J}%
_{2}$ being uniformly distributed on the sphere but pointing in opposite
directions. This distribution will now be employed for determining averages of
observables related to the angular momenta of the two particles.

\begin{figure}[tb]
\includegraphics[height=1.3in,width=3.in]{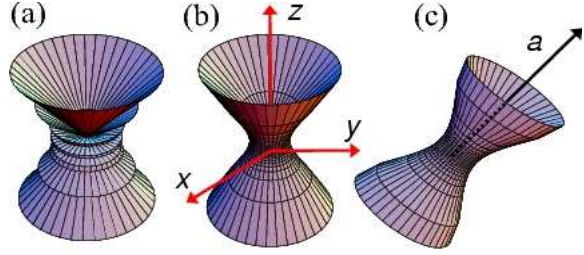}
\caption{Normalized angular distribution for a single particle in
configuration space. (a) \emph{Quantum} distribution (spherical
harmonic $|Y_{JM}(\theta,\phi)|^2$). (b) \emph{Classical}
distribution $\rho_{z_0}(\theta,\phi)$ of Eq. (\ref{7}). (c)
Classical distribution $\rho_{a_0}$ corresponding to a fixed value
of $J_a$ (here $\theta_a=\pi/4$). The angular momentum and the
projection on the $z$ [(a)-(b)] or $a$ [(c)] axis is the same for
the 3 plots
($J/\eta=40$, with $\eta=\hbar$, and $M/J=5/8$).}%
\label{f1}%
\end{figure}

\section{Statistical violation of the Bell inequalities}

Let us assume two types of detectors yielding outcomes related to the angular
momenta of the particles. The first type gives a 'sharp' (S) measurement of
$J_{1a}$ only if $J_{1a}$ is an integer multiple of some elementary gauge
$\eta$, and gives $0$ elsewhere. This detection can be represented by the
phase-space quantity%
\begin{equation}
S_{a}(\Omega_{1}) =J_{1a}(\Omega_{1})\textrm{ if }\Omega_{1}\in\Omega
_{1k},\label{11}\\
S_{a}(\Omega_{1}) =0\textrm{ elsewhere,}%
\end{equation}
where $\Omega_{1k}$ are the parts of phase space where $J_{1a}=k\eta$
compatible with a detection (see Fig.\ 2(a)). The second detector gives a
'direct' (D) measurement of $J_{2b}$ (the projection of $\mathbf{J}_{2}$ on an
axis $b$). The corresponding phase-space function is%
\begin{equation}
D_{b}(\Omega_{2})=\mathbf{J}_{2}\cdot\mathbf{b.} \label{14}%
\end{equation}
In classical mechanics there is no natural unit for quantities having the
dimension of an action, so $J$ and $\eta$ can be expressed in terms of
arbitrary units, and any physical result will depend only on the ratio
$J/\eta$. We will assume for definiteness that $\eta$ is chosen so that the
extremal values $\pm J$ can be reached. $J/\eta$ must hence be either an
integer or a half-integer, the extremal values in dimensionless units being
given by $\pm L\equiv\pm J/\eta$. For example if $\eta=2J$, the measurement
can only yield the extremal values $L=\pm1/2$ ($\eta=J$ allows to measure $\pm
L=\pm1$ and 0, $\eta=2J/3$ allows $\pm L=\pm3/2$ and $\pm(L-1)=\pm1/2$ etc.).
Note that the particle label $1$ or $2$ can be attached to the detectors:
indeed, we will call '1' the particle detected by S and '2'\ the particle
detected by D.

The classical average $E(a,b)=\left\langle S_{a}D_{b}\right\rangle $ for joint
measurements over the statistical distribution $\rho$ can be computed from%
\begin{equation}
E(a,b)=\int S_{a}(\Omega_{1})D_{b}(\Omega_{2})\rho(\Omega_{1},\Omega
_{2})d\Omega_{1}d\Omega_{2} \label{18}%
\end{equation}
with Eqs. (\ref{9}), (\ref{11}) and (\ref{14}). Given the characteristics
(\ref{11}) of the S detection, Eq. (\ref{18}) is actually a discrete sum over
the parts of phase-space $\Omega_{1k}$ leading to the detection of $k\eta$;
this can be written by including a delta function under the integral. Eq.
(\ref{9}) imposes $\theta_{2}=\pi-\theta_{1}$ and $\phi_{2}=\pi+\phi_{1},$ and
Eq. (\ref{18}) becomes
\begin{equation}
 E(a,b)=\frac{1}{2}\sum_{k=-L}^{k=L}\int\left[  L\cos\theta_{1}\right]
\delta\left(  L\cos\theta_{1}-k\right)  \left[  -L\cos\theta_{1}\cos
(\theta_{b}-\theta_{a})\right]  \sin\theta_{1}d\theta_{1}, \label{40}%
\end{equation}
where we have chosen the $z$ axis to coincide with $a$ to take advantage of
the axial symmetry imposed by $S_{a}$ (here the limiting procedure in the
delta function is understated). The $\frac{1}{2} $ prefactor is the only
nontrivial normalisation factor (coming from the integration over $\theta_{1}
$). We obtain the average as
\begin{equation}
E(a,b)=-\frac{1}{6}(L+1)(2L+1)\cos(\theta_{b}-\theta_{a}), \label{20}%
\end{equation}
which as expected depends solely on the ratio $J/\eta\equiv L$.

The correlation function employed in Bell's inequality can be obtained in the
standard (or CHSH) form \cite{CHSH69,bell2}. We choose 4 axes $a$, $b\mathbf{,} $
$a^{\prime}$, $b^{\prime}$ (we can assume an S detector is placed along ${a}$
and $a^{\prime}$, and a D detector along ${b}$ and ${b}^{\prime}$) and
determine the average values for each of the 4 possible combinations involving
an S and a D detector. The correlation function $C$ relating the average
values obtained for different orientation of the detectors' axes is%
\begin{equation}
C(a,b,a^{\prime},b^{\prime})=\left(  \left|  E(a,b)-E(a,b^{\prime})\right|
+\left|  E(a^{\prime},b)+E(a^{\prime},b^{\prime})\right|  \right)  (L)^{-2}
\label{23}%
\end{equation}
where we have divided by $L^{2}$ to obtain the CHSH function in the standard
form characterized by values bounded by $\pm1$. Here the detected values obey
the conditions $\left|  S/L\right|  \leq1$ and $\left|  D/L\right|  \leq1$, so
that the usual derivation of the Bell inequalities would lead to%
\begin{equation}
C(a,b,a^{\prime},b^{\prime})\leq2. \label{25}%
\end{equation}
By replacing Eq. (\ref{20}) in Eq. (\ref{23}), it can be seen that for
$L=\frac{1}{2}$, $1$ and $\frac{3}{2}$, there are several choices of the axes
that lead to $C(a,b,a^{\prime},b^{\prime})>2$. The maximal value of the
correlation function corresponds to $C(0,\frac{\pi}{4},\frac{\pi}{2}%
,\frac{3\pi}{4})=4\sqrt{2}$ and $2\sqrt{2}$ for $L=\frac{1}{2}\ $and $1$
respectively \footnote{The reader familiar with the Bell inequalities for the
quantum measurement of $J_{1a}$ and $J_{2b}$ will recognize the similarity of
Eq. (\ref{20}) with the quantum expectation value; the only difference is that
the quantum expectation value is normalized respective to the number of
possible outcomes ($2L+1$) whereas here the normalization \ is relative to
classical phase-space (namely the length $2L$ of the measurement axis).}.

\begin{figure}[tb]
\includegraphics[height=2in,width=3.9in]{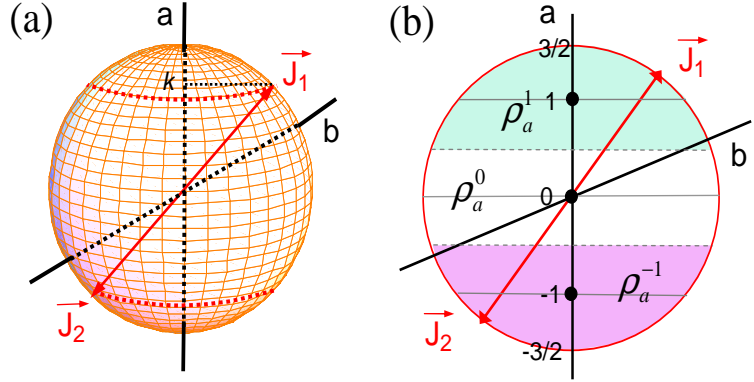}\caption{Setups for the
first (a) and second (b) examples investigated in this work. In (a) an S
detector is placed along the $a$ axis and a D detector along $b$. The angular
momenta, originally distributed on the sphere, are constrained to move on the
rings (red dotted) corresponding to fixed values of the projection on $a$. (b)
shows the $(\widehat{z,y})$ plane of the angular momentum sphere for the $L=1$ case (hence $J=3/2$); the 3
zones correspond to the projections of the spherical zones $\rho_{a}^{-1}$, $\rho_{a}^{0}$ and $\rho_{a}^{+1}$. If $J_{a}$ is measured, the presence of $\mathbf{J}$ in one of these zones yields the outcomes $J_{a}=-1,0,1$ with unit probability.  If $J_{b}$ is measured and $\mathbf{J} \in \rho_{a}^{k}$ \emph{any} of the outcomes $J_{b}=-1,0,1$ can be obtained with probabilities depending on the distribution $\rho_{a}^{k}$.}%
\label{f2}%
\end{figure}

The violation of the Bell inequality is due to the fact that we are only
including in the statistics the measurements for which \emph{both} the S and
the D detectors click. But when an S-measurement is made along the two
different orientations $a$ and $a^{\prime}$ that enter the correlation
function, different and mutually exclusive parts of phase-space are selected,
so that the different events
\begin{equation}
\{S_{1a},D_{2b}\}, \{S_{1a^{\prime}},D_{2b}\}, \{S_{1a},D_{2b^{\prime}}\},
\textrm{ and } \{S_{1a^{\prime}},D_{2b^{\prime}}\} \label{31c}%
\end{equation}
are not supported by a common phase-space distribution.\ As a consequence the
quantity
\begin{equation}
\int S_{a}(\Omega_{1})D_{b}(\Omega_{2})S_{a^{\prime}}(\Omega_{1})D_{b^{\prime
}}(\Omega_{2})\rho(\Omega_{1,}\Omega_{2})d\Omega_{1}d\Omega_{2} \label{32}%
\end{equation}
describing the average of simultaneous measurements along the 4 axes becomes
undefined. However, as we mentioned above, the existence of the joint
probability distribution in the integrand of Eq. (\ref{32}), or equivalently
\cite{accardi}, of a common distribution for the events (\ref{31c}) is a
necessary ingredient in the derivation of Bell's theorem, thereby explaining
the violation of the inequalities. It is noteworthy that if one includes the
\emph{entire} phase-space in the average (\ref{18}) instead of the parts of
phase-space corresponding to the double-click events, then Eq. (\ref{32})
becomes well-defined. It can then be shown that $E(a,b)$ and $C(a,b,a^{\prime
},b^{\prime})$ should be multiplied by the fraction of phase-space yielding
the double click measurements\footnote{Here this part of phase-space is
infinitesimal, since for the sake of mathematical simplicity we have modeled
the S-detection by a delta function. If we replace the delta functions on the
angular momentum sphere by narrow rings and spherical caps having a finite
surface, the fraction of phase-space leading to double-click events becomes
finite, and the reasoning as well as the conclusions reached with the delta
function modeling hold (although the computations need to be made numerically)
\cite{matz-p}.}: as a result Bell's inequality \emph{would not be
violated}.\ From the standpoint of classical mechanics, the objection
regarding the necessity of including the entire phase-space makes sense, since
one can envisage in principle a particle analyzer able to detect the particles
that have not been included in the double-click statistics. The quantum analog
of this problem is the well-known detection loophole, pending on the
experimental tests of Bell's inequalities \cite{percival,santos}.

\section{Dynamically induced violation of the Bell inequalities}

Our second setting goes further into the violation of Bell's inequalities by
postulating a model involving a local probabilistic interaction during the measurement between
the detector and the particle being measured: we then obtain a violation of
the inequality for the entire ensemble of particles. Let us take two identical
detectors $T_{1}$ and $T_{2}$ that give as only output the integer or
half-integer values $k=L,L-1,...-L$ of the projection $J_{1a}$ and $J_{2b} $
of the angular momenta of the particles. We choose here $L=J/\eta-1/2$, from
which it follows that the maximal readout $L$ is smaller than $J$; for
notational simplicity we put $\eta=1$ (so $J$, rather than $J/\eta$ takes
integer or half integer values). We further assume that there is an
interaction between $T_{1}$ and particle $1$ (and between $T_{2}$ and particle
$2$) affecting the angular momentum of the particle so that the transition
$J_{1a}\rightarrow k$ is a physical process due to the measurement.

We impose the following constraints on this process (which only involves a
\emph{single} particle and its measuring apparatus, hence we drop the indices
labeling the particles).
\begin{enumerate}
\item There are distributions $\rho_{a}^{k}$ such that if $\mathbf{J}\in
\rho_{a}^{k}$%
\begin{equation}
P_{k^{\prime}}^{T_{a}}(\mathbf{J}\in\rho_{a}^{k})=\delta_{kk^{\prime}}.
\label{51}%
\end{equation}
This means that if $T_{a}$ is measured and we obtain $k$ then we know that previous to the measurement
$\mathbf{J}\in\rho_{a}^{k}$ with unit probability.

\item Let $\left\langle J_{b}\right\rangle _{\rho_{a}^{k}}$ be the phase-space
average of $J_{b}$ over the distribution $\rho_{a}^{k}$, where the directions
$b$ and $a$ are assumed to be different. If $T_{b}$ is measured and
$\mathbf{J}\in\rho_{a}^{k}$, \emph{any} outcome $k^{\prime}$ can be obtained
with a \emph{non-vanishing} probability $P_{k^{\prime}}^{T_{b}}(\mathbf{J}\in\rho
_{a}^{k}).$ Our main assumption is that averaging over $T_{b}$ gives the
phase-space average of $J_{b}$, i.e. the interaction vanishes on
average.\ This constraint takes the form%
\begin{equation}
\left\langle T_{b}\right\rangle _{\rho_{a}^{k}}=\sum_{k^{\prime}=-L}%
^{L}k^{\prime}P_{k^{\prime}}^{T_{b}}(\mathbf{J}\in\rho_{a}^{k})=\left\langle
J_{b}\right\rangle _{\rho_{a}^{k}}. \label{50}%
\end{equation}
Eq. (\ref{50}) also holds if $b=a$ but then only $P_{k}^{T_{a}}=1$ is
non-vanishing hence%
\begin{equation}
\left\langle T_{a}\right\rangle _{\rho_{a}^{k}}=\left\langle J_{a}%
\right\rangle _{\rho_{a}^{k}}=k. \label{50b}%
\end{equation}

\end{enumerate}

We will not be interested here in putting forward specific models of the
interaction yielding such probabilities; it will suffice for our purpose that
a set of numbers $P_{k}$ verifying Eq. (\ref{50}) and obeying $\sum_{k}%
P_{k}=1$ can be obtained. We need to specify however the distributions obeying
Eq. (\ref{51}). It is convenient to specify $\rho_{a}^{k}$ in terms of the distribution of
$\mathbf{J}$ on the angular momentum sphere: it can then easily be seen that Eq. (\ref{50b}) is realized if
$\rho_{a}^{k}$ is taken to be the ring centered on the $a$ axis and bounded by
$k-1/2<J_{a}<k+1/2$ (see Fig. 2(b)). Then a measurement of $T_{a}$ will yield the outcome $k$
with unit probability:{%
\begin{equation}
T_{a}=k\textrm{ if }k-1/2<J_{a}<k+1/2. \label{49b}%
\end{equation}
One can of course envisage a distribution }$\rho$ obtained by combining the
elementary ensembles $\rho_{a}^{k}$. In particular the uniform distribution on
the sphere $\rho_{\Sigma}$ is the sum of the $2L+1$ spherical rings $\rho_{a}^{k}$,
\begin{equation}
\rho_{\Sigma}=\sum_{k}\frac{\rho_{a}^{k}}{2L+1} \label{z1}%
\end{equation}
and therefore if $T_{a}$ is measured the probability of finding a given value
$k$ is $P=${$1/(2L+1)$. Inversely the obtention of the given outcome $k$ is
correlated with $\mathbf{J}\in\rho_{a}^{k}$ previous to the measurement. With
$\rho_{a}^{k}$ defined in this way [Eq. (\ref{49b})], $\left\langle
J_{b}\right\rangle _{\rho_{a}^{k}}$ is computed straightforwardly and Eq.
(\ref{50}) becomes%
\begin{equation}
\left\langle T_{b}\right\rangle _{\rho_{a}^{k}}=k\cos(\theta_{b}-\theta_{a});
\label{z2}%
\end{equation}
we see again that for correlations involving averages, the knowledge of the
individual probabilities $P_{k}^{T_{b}}$ is not necessary. Note however that
for the particular case $J=1$ (i.e., $L=1/2$) the constraints (\ref{50}%
)-(\ref{49b}) as well as the normalization of the probabilities impose the
values of the $P_{k}^{T_{b}}$ irrespective of any precise physical process:
indeed $k$ can only take the values $\pm-1/2$ from which it follows that%
\begin{equation}
P_{\pm}^{T_{b}}=\frac{1}{2}\pm\left\langle J_{a}\right\rangle _{\rho_{a}^{k}%
}=\frac{1}{2}\pm k\cos\left(  \theta_{b}-\theta_{a}\right)  . \label{54b}%
\end{equation}

Let us now go back to the 2-particle problem, assuming the initial phase-space
density $\rho$ given by Eq. (\ref{9}). The expectation value
$E(a,b)=\left\langle T_{1a}T_{2b}\right\rangle $ is computed from the general
formula
\begin{equation}
E(a,b)=\sum_{k,k^{\prime}=-L}^{L}kk^{\prime}P(T_{2b}=k^{\prime}\cap T_{1a}=k)
\label{e90}%
\end{equation}
where $k$ and $k^{\prime}$ run on the possible outcomes. The probabilities of
obtaining $T_{1a}=k$ and $T_{2b}=k^{\prime}$ are obtained in the following
way. Using
\begin{equation}
P(T_{2b}=k^{\prime}\cap T_{1a}=k)=P(T_{1a}=k)P(T_{2b}=k^{\prime}|T_{1a}=k)
\label{e92}%
\end{equation}
we first determine $P(T_{1a}=k)$ by remarking that the initial distribution
$\rho$ corresponds to $\mathbf{J}_{1}$ being uniformly distributed on the
sphere.\ According to the results of the preceding paragraph, with the sphere
being cut into $2L+1$ equiprobable zones $\rho_{a}^{k}$ [see Eq. (\ref{z1})],
we have $P(T_{1a}=k)=1/(2L+1)$. We also know that an outcome $T_{1a}=k$
corresponds to\ $\mathbf{J}_{1}\in\rho_{a}^{k}$ [Eq. (\ref{51})]. From the
conservation of the total angular momentum, we infer that particle 2 must lie
in the zone\ $\rho_{a}^{-k}$ defined by $k-1/2<-J_{2a}<k+1/2$ [Eq.
(\ref{49b})]; indeed if $T_{2a}$ were to be measured we would be assured of
finding $T_{2a}=-T_{1a}=-k$. Hence the conditional probability appearing in
Eq. (\ref{e92}) is given by%
\begin{equation}
P(T_{2b}=k^{\prime}|T_{1a}=k)=P_{k^{\prime}}^{T_{b}}(\mathbf{J}_{2}\in\rho
_{a}^{-k}) \label{z3}%
\end{equation}
where $P_{k^{\prime}}^{T_{b}}$ was defined in Eq. (\ref{50}). The sum over
$k^{\prime}$ in Eq. (\ref{e90}) thus verifies Eq. (\ref{50}) and having in
mind Eq. (\ref{z2}), the expectation value becomes
\begin{equation}
E(a,b)=\sum_{k=-L}^{L}\frac{-k^{2}}{2L+1}\cos(\theta_{b}-\theta_{a}%
)=\frac{-L(L+1)}{3}\cos(\theta_{b}-\theta_{a}). \label{55}%
\end{equation}
The correlation function is again given by Eq. (\ref{23}), since the maximum
value detected by a T measurement is $L,$ not $J$. The result given by Eq.
(\ref{55}) is familiar from quantum mechanics -- it violates Bell's inequality
for $L=1/2$ with a maximal violation for $C(0,\frac{\pi}{4},\frac{\pi}%
{2},\frac{3\pi}{4})=2\sqrt{2}$. As noted for the single particle case, the
derivation of $E(a,b)$ does not depend in any way on the individual values of
the probabilities $P_{k^{\prime}}^{T_{b}}$ {but only }on the condition
(\ref{50}) regarding the particle-measurement interaction. Note that by Bayes'
theorem, it is of course equivalent to compute $P(T_{2b}=k^{\prime}\cap
T_{1a}=k)$ from $P(T_{2b}=k^{\prime})P(T_{1a}=k|T_{2b}=k^{\prime})$, ie by assuming
that $T_{2b}=k^{\prime}$ is known first.

The violation of the Bell inequalities is due to the conjunction of two
ingredients.\ The first, represented by the constraints (\ref{51}%
)-(\ref{50b}), is relative to a \emph{single} particle and its interaction
with the measurement apparatus. The second is the conservation of the angular
momentum \emph{on average}. Interestingly the first ingredient is the one that
contradicts the assumptions made in the derivation of Bell's theorem.\ The
reason is that Eqs. (\ref{51})-(\ref{50b}) are incompatible with the
introduction of elementary probability functions $p_{k}^{T_{b}}(\Omega)$ such
that%
\begin{equation}
P_{k^{\prime}}^{T_{b}}(\mathbf{J}\in\rho_{a}^{k})=\int p_{k^{\prime}}^{T_{b}%
}(\Omega)\rho_{a}^{k}(\Omega)d\Omega; \label{z5}%
\end{equation}
indeed, such probability functions would need to depend on the ensemble,
giving rise to functions of the type $p_{k}^{T_{b}}(\Omega;\rho_{a}^{k})$.
This is shown for the case $L=1/2$ in Appendix B. With this point in
mind, one can expand Eq. (\ref{e90}) (with Eqs. (\ref{e92}), (\ref{z1}) and
(\ref{z3})) as%
\begin{equation}
E(a,b)=\int\sum_{k}kp_{k}^{T_{a}}(\Omega_{1};\rho_{\Sigma})\rho_{\Sigma
}(\Omega_{1})d\Omega_{1}\int B(\Omega_{2},k)\rho_{a}^{-k}(\Omega_{2}%
)d\Omega_{2} \label{57}%
\end{equation}
with%
\begin{equation}
B(\Omega_{2},k)\equiv\sum_{k^{\prime}}k^{\prime}p_{k^{\prime}}^{T_{b}}%
(\Omega_{2};\rho_{a}^{-k}).
\end{equation}
The dependence of $B$ on $k$ is the crucial property allowing to violate
Bell's inequality (whereas the dependence of $\rho(\Omega_{2})$ on $k$ in Eq.
(\ref{57}) by itself can be absorbed in the initial correlation $\delta
(\mathbf{J}_{1}+\mathbf{J}_{2})$ provided $k=k(\Omega_{1})$). The dependence
of $B$ on $k$ has nothing to do with non-locality or action at a distance. It
is a simple consequence of the logical inference characterizing the
conditional probability (\ref{e92}) given the characteristics of the single
particle interaction with the measuring apparatus, namely the fact that the
model allows only specific types of correlations: in the \emph{single}
particle problem one can only correlate a given outcome with a specific
distribution -- this happens when the distribution is symmetric relative to the detector's
axis [Eq. (\ref{51})]; in the \emph{two} particle problem the single particle property just mentioned makes only possible the correlation of $\mathbf{J}_{2}$ as a function of $\mathbf{J}_{1}$ in terms of the
\emph{ensembles} to which they belong, not in terms of their \emph{individual}
positions.\ This is consistent with the fact that the knowledge of the individual position
of $\mathbf{J}$ is meaningless to compute the observed probabilities, as even
the elementary probabilities must depend on the ensemble to which the angular
momentum belongs\footnote{It would be of course extremely valuable to understand
what kind of physical processes are compatible with this type of behaviour (for example the value
of the angular momentum in this case could represent some time average of an underlying stochastic process, or a space average of a field-like quantity distributed all over the ensemble).}.

Note finally that would $B$ not depend on $k$ (and the elementary
probabilities on the ensembles), Eq. (\ref{57})\ would turn into%
\begin{equation}
E^{BT}(a,b)=\int A(\Omega_{1})B(\Omega_{2})\rho(\Omega_{1},\Omega_{2}%
)d\Omega_{1}d\Omega_{2} \label{62}%
\end{equation}
where
\begin{equation}
A(\Omega_{1})\equiv\sum_{k}kp_{k}^{T_{a}}(\Omega_{1})\textrm{ \ \ }B(\Omega
_{2})\equiv\sum_{k^{\prime}}k^{\prime}p_{k^{\prime}}^{T_{b}}(\Omega_{2}),
\label{64}%
\end{equation}
thereby yielding the familiar form taken by the expectation value in the
derivation of Bell's theorem. In the \emph{deterministic} case considered by
Bell \cite{bell2} the functions $p_{k}^{T_{a}}$ and $p_{k^{\prime}}^{T_{b}}$
are either $0$ or $1$ depending on the individual position of $\mathbf{J}_{1}$
(resp. $\mathbf{J}_{2}$). This implies that $k=k(\Omega_{1})$, ie a given outcome depends
on the position of $\mathbf{J}_{1}$ on the angular momentum sphere, and
$\rho_{a}^{-k}(\Omega_{2})=\rho(\Omega_{2}|\mathbf{J}_{1})$ does not depend on
$k$ or $a$ but on $\mathbf{J}_{1}=-\mathbf{J}_{2}$ (hence the inclusion of the
term $\delta(\mathbf{J}_{1}+\mathbf{J}_{2})$ in the definition of $\rho
(\Omega_{1},\Omega_{2})$). Conversely one may \emph{assume} $\rho(\Omega
_{2}|k)=\rho(\Omega_{2}|\Omega_{1})$ in Eq. (\ref{57}) with $p_{a}^{k}%
(\Omega_{1})$ and $p_{b}^{k^{\prime}}(\Omega_{2})$ being probability functions
different from $0$ or $1$; then $A$ and $B$ defined in Eq. (\ref{64}) are not
the observed outcomes but their averages, and $E^{BT}(a,b)$ is the expectation
corresponding to the \emph{stochastic} case considered by Bell. Bell's
stochastic case correlates the individual positions of $\mathbf{J}_{1}$ and
$\mathbf{J}_{2}\ $to possible outcomes with definite probabilities. In the
present model the random interaction forbids to make the correspondence
between a given position of the angular momenta and a definite outcome;
instead the correspondence is between a definite outcome and a given ensemble
describing the positions of the angular momenta compatible with the outcome
(of course if the former correspondence is satisfied, so is the latter, but the converse is
not true). In the latter case, the structure of the expectation value
(\ref{57}) does not allow to define a term of the type given by Eq. (\ref{32})
whereby a single distribution can account for several simultaneous joint
measurements. It appears indeed that the ensemble dependency exhibited by the
present model is a necessary feature in order to produce non-commuting
measurements \cite{matz-p}. In this sense the present model can be seen as a
classical analogue of the quantum measurement of two non-commuting observables
(such as $J_{1a}$ and $J_{1a^{\prime}}$) applied to correlations between two
particles as originally considered by EPR \cite{EPR}.

\section{Conclusion}

The present results show that averages obtained with 2-particle classical
distributions constructed to be the analogs of quantum mechanical eigenstates
can violate Bell's inequalities. The violation does not involve nonlocality
but statistical or dynamical processes that impede the existence of joint
probability distributions or the correlation between individual values of the
variables as required by Bell's theorem. Possible implications on the role of the
Bell-CHSH argument as a marker of quantum nonlocality, which has recently been
criticized \cite{unruh}, will be examined elsewhere \cite{matz-p}.

\appendix
\section{}
The scheme we are employing to contruct the classical distributions rests on
the well-known analogy between the classical Poisson brackets and the quantum
commutation relations in the density matrix formalism. Let $\hat{G}$ be an
operator and $\left\vert \psi_{g}\right\rangle $ an eigenstate with eigenvalue
$g.$ Then the pure-state density matrix $\hat{\rho}_{g}\equiv\left\vert
\psi_{g}\right\rangle \left\langle \psi_{g}\right\vert $ verifies $[\hat{\rho
}_{g},\hat{G}]=0$ and $\hat{G}\hat{\rho}_{g}=g\hat{\rho}_{g}$. In classical
mechanics the Poisson bracket of two phase space quantities $u(q,p)$ and
$v(q,p)$ is a canonical invariant defined by \cite{goldstein}%
\begin{equation}
\{u,v\}=\frac{\partial u}{\partial q}\frac{\partial v}{\partial p}%
-\frac{\partial u}{\partial p}\frac{\partial v}{\partial q}.
\end{equation}
Let $\rho(q,p)$ be the phase-space distribution and $G(q,p)$ be a function
such that $\{\rho,G\}=0.$ This means that $\rho$ is invariant relative to the
canonical tranformation generated by $G$, ie%
\begin{equation}
\{\rho,G\}\delta Q_{G}=\delta\rho=0,
\end{equation}
where $Q_{G}$ is canonically conjugate to $G$, which is a constant of the
motion. Then every point of the distribution $\rho$ will be characterized by
the constant value taken by $G$, denoted $g$. If this is the only constraint
imposed on the distribution, $\rho(q,p)$ will take the form (up to a
normalization constant)%
\begin{equation}
\rho(q,p)=\delta(G(q,p)-g).
\end{equation}
In configuration space, the distribution $\rho(q)$ is obtained by integrating
over the values of the momentum compatible with a given $q,$%
\begin{equation}
\rho(q)=\int\rho(q,p)dp=\int\frac{\delta(p-p_{i})}{\left.  \frac{\partial
G}{\partial p}\right\vert _{p_{i}}}dp
\end{equation}
where $p_{i}$ is the root (assumed to be unique, else a sum is in order) of the argument of the delta
function. Integrating yields%
\begin{equation}
\rho(q)=\left.  \frac{\partial p}{\partial G}\right\vert _{p_{i}}=\left.
\frac{\partial^{2}S}{\partial q\partial G}\right\vert _{p_{i}}%
\end{equation}
where $S(q,G)$ is the classical action. The configuration space density is
therefore the amplitude of the quantum density matrix element $\left\langle
q\right\vert \hat{\rho}_{g}\left\vert q\right\rangle $ in the semiclassical approximation.

\section{}
We show that the detection model for a single particle given in Sec. 4 is
inconsistent with probability functions defined by Eq. (\ref{z5}) in the
$L=1/2$ case (the one violating the Bell inequalities). Take Eq. (\ref{z5})
with $a=b$ and $k,k^{\prime}=+1/2$,
\begin{equation}
P_{+}^{T_{b}}(\mathbf{J}\in\rho_{b}^{+})=\int p_{+}^{T_{b}}(\Omega)\rho
_{b}^{+}(\Omega)d\Omega=1. \label{a1}%
\end{equation}
Particularizing the general formula (\ref{49b}) to the case $L=1/2$,
$\rho_{b}^{+}$ is the positive hemisphere of the unit sphere (since $J=1$)
centered on the $b$ axis. The result on the right handside follows from Eq.
(\ref{54b}). Eq. (\ref{a1}) implies that $p_{+}^{T_{b}}(\Omega)=1$ for
$\mathbf{J}\in\rho_{b}^{+}$ and consequently $p_{-}^{T_{b}}(\Omega)=0$.
Conversely since $P_{+}^{T_{b}}(\mathbf{J}\in\rho_{b}^{-})=0$, we must have
$p_{+}^{T_{b}}(\Omega)=0$ and $p_{-}^{T_{b}}(\Omega)=1$ when $\mathbf{J}%
\in\rho_{b}^{-}$. Now assume that the distribution is instead $\rho_{a}^{+}$
with $a$ different from the $b$ axis. Then according to our model [Eq.
(\ref{54b})] we should have%
\begin{equation}
P_{+}^{T_{b}}(\mathbf{J}\in\rho_{a}^{+})=\int p_{+}^{T_{b}}(\Omega)\rho
_{a}^{+}(\Omega)d\Omega=\cos^{2}\frac{\theta_{b}-\theta_{a}}{2}. \label{a2}%
\end{equation}
Noting that $\rho_{a}^{+},$ the positive hemisphere centered on $a$, is
actually composed of two parts, $\rho_{a}^{+}\cap\rho_{b}^{+}$ and $\rho
_{a}^{+}\cap\rho_{b}^{-}$ we can write%
\begin{equation}
P_{+}^{T_{b}}(\mathbf{J}\in\rho_{a}^{+})=\int_{\rho_{a}^{+}\cap\rho_{b}^{+}%
}p_{+}^{T_{b}}(\Omega)\rho_{a}^{+}(\Omega)d\Omega+\int_{\rho_{a}^{+}\cap
\rho_{b}^{-}}p_{+}^{T_{b}}(\Omega)\rho_{a}^{+}(\Omega)d\Omega.
\end{equation}
\ But we have seen that $p_{+}^{T_{b}}=1$ for $\mathbf{J}\in\rho_{b}^{+}$ and
$p_{+}^{T_{b}}(\Omega)=0$ for $\mathbf{J}\in\rho_{b}^{-}$, hence%
\begin{equation}
P_{+}^{T_{b}}(\mathbf{J}\in\rho_{a}^{+})=\int_{\rho_{a}^{+}\cap\rho_{b}^{+}%
}\rho_{a}^{+}(\Omega)d\Omega=1-\frac{\theta_{b}-\theta_{a}}{\pi},
\end{equation}
which contradicts Eq. (\ref{a2}). Hence probability functions obeying Eq.
(\ref{z5}) do not exist, and Eq. (\ref{z5}) should be replaced by%
\begin{equation}
P_{k^{\prime}}^{T_{b}}(\mathbf{J}\in\rho_{a}^{k})=\int p_{k^{\prime}}^{T_{b}%
}(\Omega;\rho_{a}^{k})\rho_{a}^{k}(\Omega)d\Omega
\end{equation}
where the notation $p_{k^{\prime}}^{T_{b}}(\Omega;\rho_{a}^{k})$ denotes the
dependence of the elementary probabilities on the distribution. Note also that
Eq. (\ref{z5}) \emph{does} hold if one drops the requirement that $p_{k^{\prime}%
}^{T_{b}}(\Omega)$ should represent an elementary probability: for example the
functions $p_{+}^{T_{b}}(\Omega)=J_{b}+1/2$ or $p_{+}^{T_{b}}(\Omega
)=2J_{b}H(J_{b})\ $fulfill Eq. (\ref{a2}) without depending on the
distribution, though none of these functions is contained in the interval
$[0,1]$ and are thus not probability functions. We stress that these features, which put strong
constraints on the type of admissible physical models that one could envisage, are relevant to a
\emph{single} particle and its interaction with the measurement apparatus.

\end{document}